\documentclass[]{raa}
\usepackage{graphicx}
\usepackage{dcolumn}
\usepackage{mathrsfs,mathtools}
\usepackage{amsmath,amssymb}   
\usepackage{url}
\usepackage[utf8]{inputenc}
\usepackage[T1]{fontenc}
\usepackage{mathptmx}

\usepackage{natbib}
\usepackage{color}
\usepackage{aas_macros}

\begin{document}

   \title{The Design and Implementation of a ROACH2+GPU based  Correlator  on the Tianlai Dish Array
}
   \volnopage{Vol. 000 No.0, 000--000}      
   \setcounter{page}{1}          

   \author{Chenhui Niu
      \inst{1,2,3}
   \and Qunxiong Wang
      \inst{2,4}
    \and David MacMahon
    \inst{3}  
   \and Fengquan Wu
      \inst{2}
      \and Xuelei Chen 
       \inst{2,5,6}
      \and Jixia Li
       \inst{2,5}
   \and Haijun Tian
       \inst{4}
  \and Guillaume Shippee
       \inst{3}
   \and Dan Werthimer
   	\inst{3}
   \and Xiaoping Zheng
       \inst{1}
   }

   \institute{Central China Normal University, Luoyu Road, Wuhan 430079, China 
        \and
            Key Laboratory for Computational Astrophysics, National Astronomical Observatories, Chinese Academy of Sciences, Beijing 100012, China\\
        \and
          University of California Berkeley, Campbell Hall 339, Berkeley CA 94720\\
          \and 
          China Three Gorges University, Yichang,443002, China\\
          \and
 School of Astronomy and Space Sciences, University of Chinese Academy of Sciences, Beijing 100049, China\\
 	\and
 Center of High Energy Physics, Peking University, Beijing 100871, China	\\
 Correspondence: {\tt xuelei@cosmology.bao.ac.cn, wufq@bao.ac.cn, zhxp@mail.ccnu.edu.cn}
}
\authorrunning{C.H.Niu et al.}

\titlerunning{Correlator on the Tianlai Dish Array}

\date{Received~~xxx October xx; accepted~~xx~~Feb xx}

\abstract{The digital correlator is a crucial element in a modern radio telescope. In this paper we describe a  scalable design of the 
correlator system for the Tianlai pathfinder array, which  is an experiment dedicated to test the key technologies for 
conducting 21cm intensity mapping survey.  The correlator is of the FX design, which firstly performs Fast Fourier Transform (FFT)
including Polyphase Filter Bank (PFB) computation using a Collaboration for Astronomy Signal Processing and Electronics Research (CASPER)  Reconfigurable Open Architecture Computing Hardware-2 (ROACH2) board, then computes cross-correlations  using Graphical Processing Units (GPUs). The design has been tested both in laboratory and in actual observation.\\
\keywords{instrumentation: interferometers}
}

\maketitle

\section{\label{sec:level1}Introduction}

In a radio astronomy telescope array, the correlator is implemented to obtain the short time average of signal correlations, which is 
called the (interferometric) visibilities. By performing the cross-correlation, the phase information of the radio wave is preserved, and
the sky radio intensity map can be reconstructed in the synthesis imaging 
process by using the visibilities as input data \citep{2001isra.book.....T}. The correlator thus serves a central function in 
radio telescopes.   
 	 
The Tianlai project is a 21cm intensity mapping experiment with the aim of measuring the baryon acoustic oscillation (BAO) features in 
the matter power spectrum \citep{2012IJMPS..12..256C}. Currently, two pathfinder arrays have been built
in a radio quiet site in Hongliuxia, Balikun county, Xinjiang, China.
The cylinder array includes three cylinder reflectors, which are 15 meters wide and 40 meters long oriented in the north-south direction,  
It has a total of 96 dual polarization receivers, which generates 192 input channels.  The dish array includes 
16 dishes of 6-meter aperture. Each dish has a dual polarization receiver, and a total of 32 input channels are generated. The radio signal collected
in the antenna feed is amplified by a low noise amplifier, which is then converted to optical signal and transmitted via an optical cable
to the station house located 8 km away, Afterwards, the optical signal is converted back to radio signal and sent to the receiver. The analog receiver is 
of a heterodyne type with the intermediate frequency band of 135-235 MHz. The digital backend samples this signal at a rate of 
250 Mega samples per second (MSPS).  The correlator handles the full polarizations, 
and produce both cross-correlations and auto-correlations.

A prototype system with 32 input channels based on this channel has been built and installed on the pathfinder Tianlai dish array. 
The design of this system (e.g. the employment of a network switch instead of communication within one chassis)
also allows it to be expanded to a scale which can handle the 192 receiver 
channels. The current Tianlai cylinder array uses a 192-channel digital correlator built by the  Institute of Automation of the Chinese 
Academy of Sciences, which uses FPGA boards and digital signal processors (DSP) of their own design. 
Here we introduce the 32-channel system as well as the 192-channel system design in some detail for future reference.

\section{System Design}\label{sec:1}
The visibilities of the interferometry array are usually computed in either of the following two ways:  1) the XF type that the 
time-ordered voltage signal from the different receiver channels are paired and convolved with each other to produce the 
cross-correlations. The Fourier transform
is then performed to obtain the visibilities at different frequencies; 
2) the FX type that the voltage signal of each receiver channel is first Fourier transformed to produce a spectrum,
and then each pair is cross-correlated for the same frequency.
 In either way, the result is integrated for some pre-specified duration to yield the final output. 
 As multiplying in frequency domain is equivalent to convolving in the time domain, the results from these two types of correlators are 
 identical. With modern digital technology and larger telescope arrays, the FX type is more convenient to 
 implement \citep{2016JAI.....502002P}, thus it has became more popular, and we also choose the FX type in this design. 
 
 One design of the Tianlai correlator system we considered is based on 
an Field Programmable Gate Array (FPGA) board for the data sampling and performing the fast Fourier transform (FFT). An instantaneous 
frequency spectrum is then derived from the time series data. The data from different channels are then transposed, 
i.e. re-arranged so that the data with same frequencies from different receivers are put together. 
 The transposed data is sent to an array of  Graphic Processing Units (GPUs) via a 10 Gbit/s  network switch, which computes the cross-correlations. Computationally, the cross-correlation is a multiply and accumulation (CMAC) process. The FPGA board we used 
is the CASPER ROACH2 board \footnote{ \url{https://casper.berkeley.edu/wiki/ROACH2}}, which has been widely used 
in radio astronomy projects \citep{2016JAI.....541001H}  (e.g. the Precision Array for Probing the Epoch of Re-ionization 
(PAPER) \citep{2010AJ....139.1468P} ). 
Our design is built upon the PAPER correlator model \citep{2008PASP..120.1207P}, which creates 
a flexible and scalable hybrid correlator system.
 
Our correlator consists of the F-engine, the network switch and the X-engine. The
F-engine is dedicated to perform the Fourier transform, while the
X-engine is dedicated to compute the cross-correlations. The network
switch is used to transpose the data. We use the CASPER ROACH2 board for the F-engine and 
 the NVIDIA$^{\rm TM}$ GPU board for X-engine.  
 The required number of ROACH2 boards is dependent on the number of receiver inputs.
Suppose the F-engine consists  of $M$ ROACH2 nodes, and the X-engine has $N$ GPU nodes.  Every ROACH2 node handles 
$m$-way analog radio inputs, and after performing the FFT, it outputs $m$ spectra. Each spectrum has 
$F$ frequency points. To optimize data traffic, each GPU node processes ${F}/{N}$ frequency points, and each frequency point 
includes $M\times m$ conjugate multiply computations. Therefore the ROACH node should divide the $F$ point spectrum into $N$ 
bands, and then send the specified band to the corresponding GPU node for the CMAC computation.

\begin{figure*}
\begin{center}
 \includegraphics[width=0.7\textwidth]{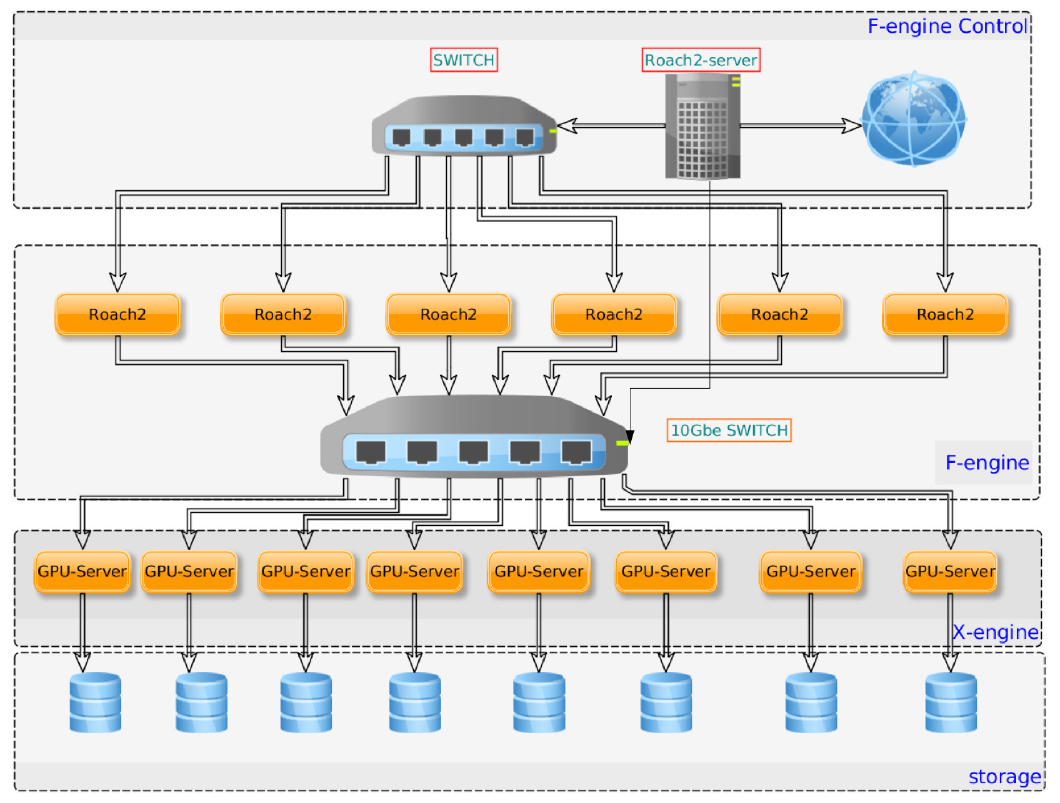}
\caption{Block diagram of the Tianlai dish array correlator.\label{fig:structure}}
\end{center}
\end{figure*}
 	  
A block diagram of the correlator structure is shown in Fig.\ref{fig:structure}. The F-engine is built with CASPER ROACH2 (R2 2012 version).  Each ROACH2 board is connected with two daughter ADC boards through 
Z-DOK+ connectors. The ADC board is the CASPER ADC 16$\times$250-8 Q2 2012 version, which 
uses 4 HMCAD1511 chips made by Hittite Microwave$^{\rm TM}$ with a total of 16 inputs \footnote{\url{https://casper.ssl.berkeley.edu/wiki/ADC16x250-8}} and 
samples 16 analog signal inputs with 8 bits at 250 Msps. The ADC input ports are Sub-Miniature-A connector, which requires analog signal of -8.5 dBm Gausian Noise. We find that good linearity can be achieved for the  input ranging from -12dBm to 6dBm. 

The ROACH nodes are controlled by a ROACH2-server, in our case it is built on a Dell$^{\rm TM}$ PowerEdge T110 with 
Intel$^{\rm TM}$ Xeon(R) CPU E31220, which can itself be controlled remotely through the Internet. 
The ROACH2 boards can be booted either by loading the kernel from the ROACH2 server (net boot) 
or from the on-board memory  chip (solo boot). A BORPH (Berkeley Operating System for ReProgrammable Hardware) 
operating system, which is a full-featured Linux operating system supporting FPGA applications\citep{So:EECS-2007-92} is run on the
ROACH2 boards, and the ROACH2 board can be controlled via an interface called the Karoo Array Telescope Control 
Protocol (KATCP)\citep{2016MNRAS.460.1664F}.  Compiled binary executable programs (bof file) can be uploaded to the 
ROACH2 boards as firmware.
 
The output data of F-engine are initially grouped by receiver channels, i.e. a block of spectra have the same receiver channel but 
different frequencies. However, only the correlations for the data of the same frequency are non-zero and need computation, and 
for this computation the spectra from different receiver channel pairs are needed by the computing unit. So the data
should be re-arranged according to their frequencies. This is done by pre-specified program in the F-engine. The data are 
packaged and then sent to different ports on the X-engine.

In the X-engine, the correlations are obtained and then sent to a hard drive server for storage. 
Our X-engine is built with the NVIDIA GTX 690 GPUs  \footnote{\url{https://www.geforce.com/hardware/desktop-gpus/geforce-gtx-690}}. 
Each GPU node has two GTX 690 units, which is equipped with two GPU cores. 
The CASPER High Availability Shared Pipeline Engine (Hashpipe) \footnote{\url{https://github.com/david-macmahon/hashpipe}} 
is used to manage the GPU  threads and data transfer within each GPU node. 
		
For the dish array, one ROACH2 board can handle the 32 inputs, and the cross-correlation can be done with 1 GPU node.
Indeed, in this case there is no need of using network switch, the data from F-engine can be sent  to the X-engine port directly. 
However,  as an option of the cylinder correlator, we also considered a design of a system with 192 receiver input channels.
In this case, the F-engines consists of 6 ROACH2 boards, and each board handles 32 inputs.  For the X-engine,  we estimate the total amount
of computation for the 192-input system is 35.6 TFLOPS, while the computing performance of each node is 4.8 TFLOPS. We thus estimate at least 8 GPU nodes (16 GPUs) are required for computation.

\subsection{F-engine}\label{sec:F-engine}

The F-engine includes 4 main function blocks: 1) the FFT block that
channelizes the digital data from ADC; 2) the Equalizer block which
truncates the data to reduce the size; 3) the Transpose block which
transpose the data; 4) the Ethernet block that sends the data to the
corresponding GPU nodes based on the frequency.
 
 As the functions and requirements 
 of our correlator are very similar to that of the PAPER experiment correlator which also used the ROACH2 system, 
 we made our design on the basis of their design\footnote{\url{https://casper.berkeley.edu/wiki/PAPER_Correlator_Manifest}}.
 The whole design can be simulated with the Matlab Simulink$^{\rm TM}$  CASPER tool set, which has a graphic user interface (GUI)
for the FPGA programming.

\begin{figure*}
 \centering
 \includegraphics[width=0.9\textwidth]{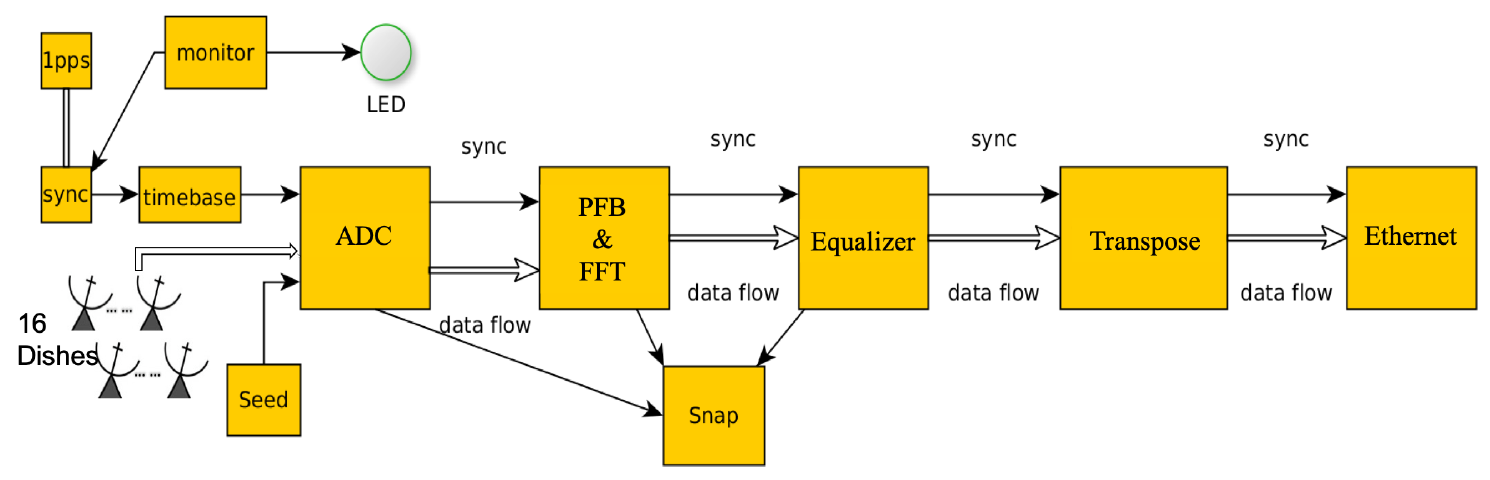}
\caption{The F-engine data flow block diagram. The data flows starts from ADC, goes though the PFB, equalizer, Transpose,  
then send via an Ethernet interface to the X-engine.    \label{fig:f-engine}}
\end{figure*}

An F-engine data flow block diagram is shown in Fig.\ref{fig:f-engine}. 
The whole system is triggered and synchronized with the 1 pulse per second (1PPS) block.  A monitor 
block is also added to watch the status of operation. A Seed block is added for testing by generating digital 
pseudo-random Gaussian noise to ADC. Additionally, a Snap block is added to
debug the ADC, PFB and Equalizer.

Each input data stream from the ADC is channelized in frequency domain with a polyphase filter bank (PFB) to minimize energy leakage. A hamming window is multiplied within each data stream. Two FFT length options are implemented in our design, which are 512 and 1024 respectively.

The FFT of the real number input will generate complex numbers, but with a Hermitian symmetry on the Fourier components, i.e. 
the negative frequency part is just the complex conjugate of the corresponding positive frequency 
component, only half of the complex numbers are independent. To avoid wasting memory and data transportation, 
the input data block of two input channels are combined as the real and imaginary components respectively to form 
a complex input, and then FFTed together \citep{1996dspp.book.....P}.
Taking two time series $x_1[n],x_2[n]$ as real part and imaginary part, the combined complex number is given as:
\begin{equation*}
y[n]=x_1[n]+j \cdot x_2[n].
\end{equation*}
Assume the Fourier pair as:
\begin{equation}
y[n] \leftrightarrow S[\nu]
\end{equation}
The Fourier transform of $x_1[n],x_2[n]$ can be recovered with the Fourier transform of y[n], i.e. 
\begin{equation}
\begin{aligned}
\mathscr{F}\{x_1[n]\} &= \frac{S_r[\nu]+S_r[-\nu]}{2}+j\cdot \frac{S_i[\nu]-S_i[-\nu]}{2}\\
\mathscr{F}\{x_2[n]\} &= \frac{S_i[\nu]+S_i[-\nu]}{2}-j\cdot \frac{S_r[\nu]-S_r[-\nu]}{2}
\end{aligned}
\end{equation}
where $S_r[\nu]$ and $S_i[\nu]$ are the real and imaginary part of the Fourier transform of y[n]. This recovery is 
implemented in the block. The output data numbers of the PFB block are 36 bit long, with the first 18 bits for the real
part and the last 18 bits the imaginary parts.  

\begin{figure*}
 \centering
 \includegraphics[width=0.7\textwidth]{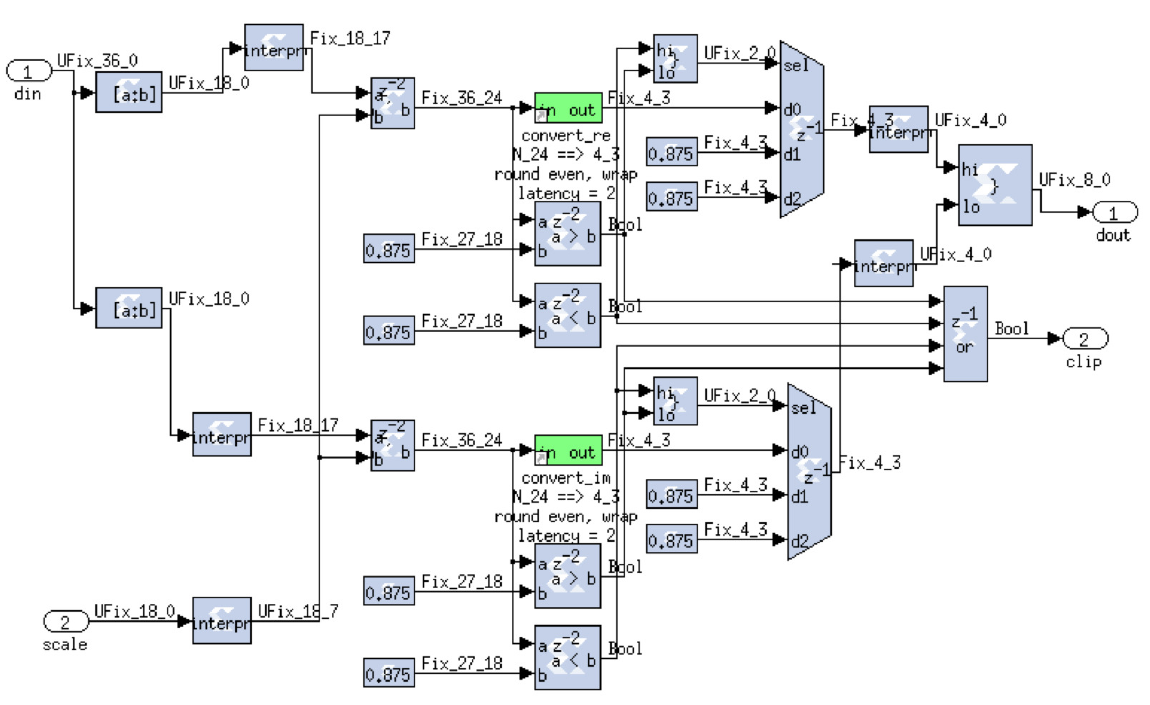}
\caption{Equalizer block. The scale factor could be varied both with different frequency and input channel.\label{fig:equalizer}}
\end{figure*}

In the pipeline, an Equalizer block follows the FFT. It serves two purposes: (1) To compensate for the varying
spectral response by multiplying the data with a frequency dependent adjustment 
factor, so as to obtain a nearly flat spectral response, and achieve good dynamical range in the digital sampling. 
(2) To reduce the amount of data to be exchanged via the network switch, the original 18 bit data 
is truncated to 4 bits by the equalizer. To preserve the maximum amount of information, the equalizer should be designed 
such that the data should most frequently fall in a suitable range. For radio astronomy, the data is usually noise-dominated, 
with occasional outliers mostly coming from radio frequency interferences (RFIs). 
For such a signal, the RMS can be estimated from the autocorrelation power levels, 
\begin{equation}
{\rm RMS}= \sqrt{\frac{P}{2N}},
 \end{equation}
where $P$ is the integrated autocorrelation value and $N$ is the number of samples in the integration. 
The equalizer is dedicated to desaturate the output while
preserving most information. The RMS of the 4-bit output falls between 2
and 3, out of the full 4-bit range of -7 to +7 \footnote{\url{https://casper.ssl.berkeley.edu/wiki/PAPER_Correlator_EQ}}. 
Considering the RFI, the equalizer is designed as follows (see Fig.\ref{fig:equalizer}). The 36 bits 
complex data is divided into 18 bits real part  and 18 bits imaginary part. It is then multiplied with the scale factor, 
which is an unsigned fix\_18\_7 number, i.e. an 18 bit long fixed point number with 7 bits after the binary point,  
and the product is a fix\_36\_24 number. The most significant bits are usually 0, except for the strong RFI. 
This number is then rounded even from the 21 bit, then the 22nd to 25th bits are selected while the other bits are discarded, 
the output is a fix\_4\_3 number, in the range of $1.001_{\rm {b}}$  to $0.111_{\rm {b}}$.
The two 4 bit real numbers are then re-packed to a 8-bit complex number and sent to the next block in the pipeline.

\begin{figure*}
 \centering
 \includegraphics[width=0.8\textwidth]{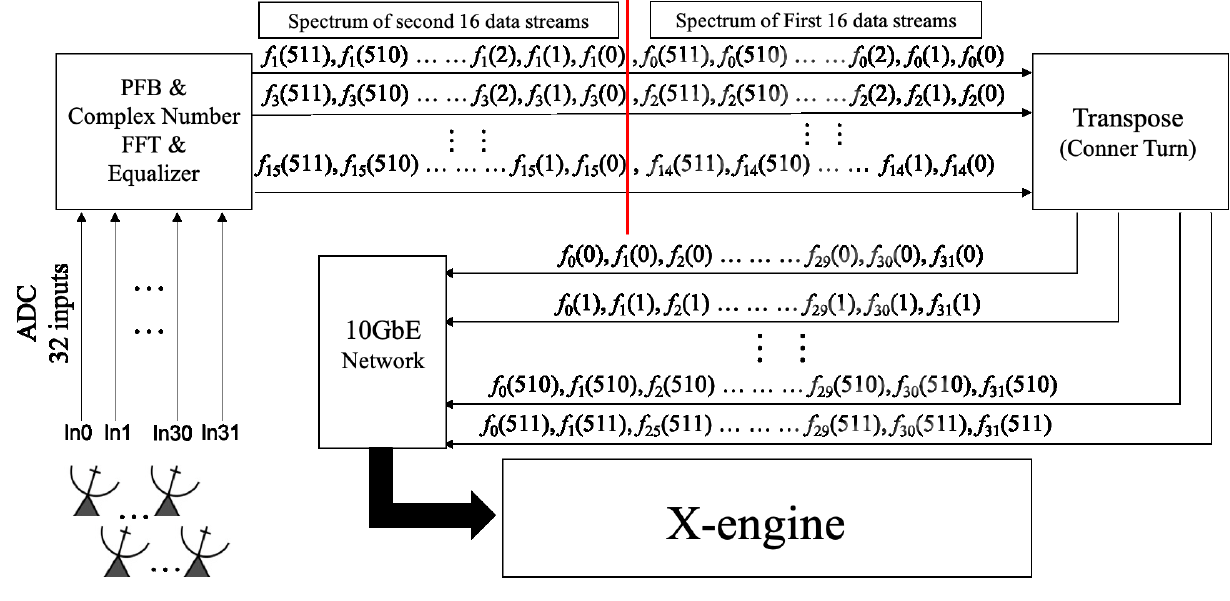}
\caption{Data Transpose from the original grouping to the transposed one.$f_i(\nu)$ is the Fourier transform of time stream, subscript i is the anntena input index that in range 0 to 31, $\nu$ is frequency channel index which range from 0  to 511 for 512 frequency bins model in this illustration. \label{fig:transepose}}
\end{figure*}

After the FFT step, the data is sent to the X-engine for cross correlation computation. 
This computation is distributed over multiple units, and each unit would need the same frequency signal from all receivers units. 
The data needs to be transposed (the so called "corner turn"), so as to arrange the data in the desired order, i.e. 
converting the  data shape from [input channel, frequency] to [frequency, input channel], as shown in Fig.\ref{fig:transepose}.

This is achieved in the F-engine by writing the data blocks consecutively to a dual-port RAM, and then read out the data not 
in the original order, but at pre-specified addresses, so that it is regrouped in the desired order. 
The regrouped data are then packaged with destination address in their headers and send to the X-engine 
via the network interface controllers (NICs) on board.  
As we have four 10GbE NICs for each ROACH2 board, the spectrum is split into 4 sub-bands, denoted by tid=0,1,2,3.

The frequency ordered data is divided into 4 blocks, so that each 
block contains 128 (for 512 frequency bins) or 256 (for 1024 frequency bins) frequency points, 
which are consecutively written into the 4 dual port RAMs.
Each FPGA processes 32 input channels from the ADCs. The 32 inputs are sampled in parallel which are written as 16 
complex numbers, with each complex number corresponding to the spectrum of two real inputs. 
 After going through the Equalizer, the length of the data is 4 bit for each real number, and 8 bit in total for the complex number.   
Within such a sub-block, the data already has 
the same frequency but different input channels, which is suitable for the use by the 
X-engine.
The reading program will read 64 bits of data 
at a time, corresponding to 16 inputs of 4 bits, so it could get the 32 inputs of the same frequency by reading in 
two numbers from the RAM. 

\begin{figure}
 \centering
 \includegraphics[width=0.35\textwidth]{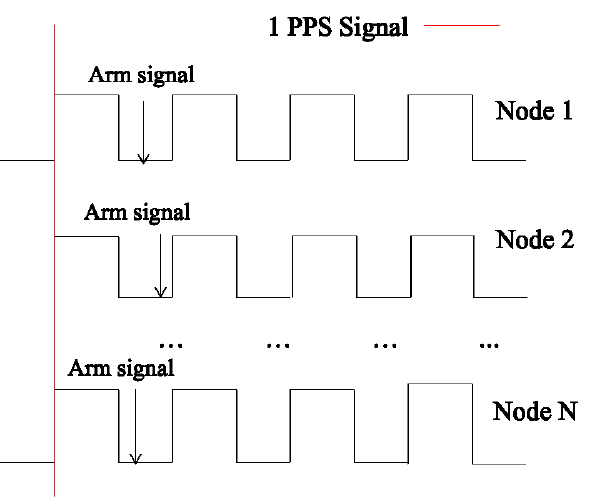}
\caption{Synchronization of the ROACH2 nodes. The red lines indicate the start trigger using 1PPS signal \label{fig:1pps}}
\end{figure}

 Synchronous ports are used in the functional block design, to ensure the synchronization of different ROACH2 nodes to the same 
 clock cycle. After all parameters such as the IP and MAC addresses are set, a synchronization signal called the "Arm signal" is 
 sent to each node. Then, all the ROACH2 nodes are initialized and waiting for an 1PPS signal to trigger the system. The 
 Arm signals at the different nodes are not required to be synchronized, but the 1PPS signals are synchronized by 
 choosing the same length of cable for each signal. After this operation, all the ROACH2 nodes will be synchronized. 
A diagram is shown in Fig.\ref{fig:1pps}.

\subsection{X-engine}\label{sec:X-engine}

When the F-engine is initialized, a trigger signal is also sent to initiate the X-engine. 
The X-engine receive the data from the F-engine in packets, which are delivered into the different computing nodes, where the 
CMAC computations are done. For one GPU node, there are two NVIDIA GTX690 cards, each containing 2 GPU cores, 
2  Intel$^{\rm TM}$ Xeon CPUs and 4 10Gbps NICs. 
These four GPU cores work separately, each processing the CMAC of 256 frequency points for the 
 32 inputs correlator,  or 32 frequency points if there are 192 inputs and 1024 frequency channels.  
 The CMAC process uses the {\tt xGPU} package \citep{2013IJHPC..27..178C}, which is written in CUDA-C and is optimized 
on GPU memory resources by specific thread tasks. 

\begin{figure}
 \centering
 \includegraphics[width=0.5\textwidth]{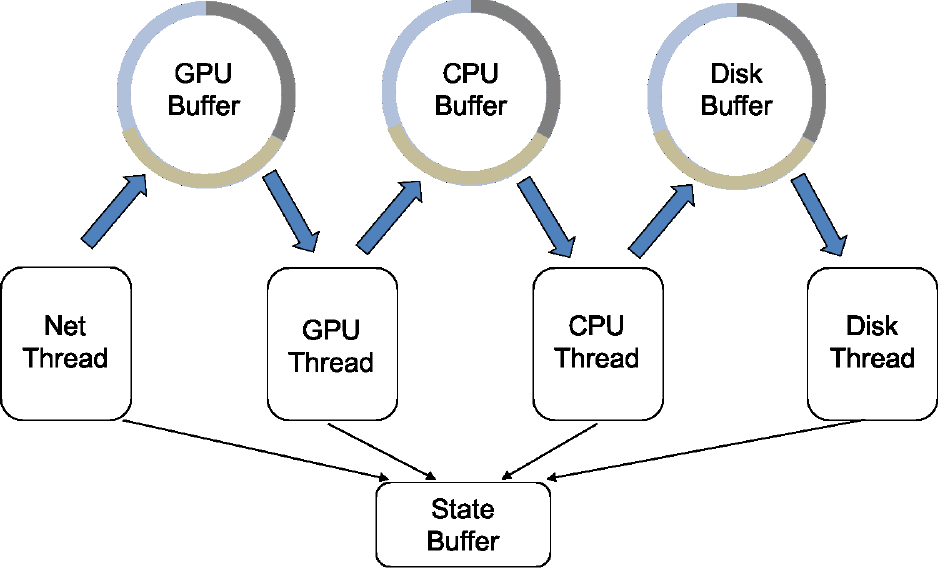}
\caption{Hashpipe thread manager diagram.\label{fig:hashpipe}}
\end{figure}

The data movement in the X-engine is managed by the Hashpipe\footnote{\url{https://github.com/david-macmahon/hashpipe}}, 
which specializes in communication between threads in both GPUs and CPUs. A sketch of the data flow in the X-engine pipeline
 is shown in Fig.\ref{fig:hashpipe}. We open 4 Hashpipe threads for each GPU node (note these are NOT the GPU threads)
 and also set up 3 ring buffers in the memory 
 for each GPU node. Each ring buffer is divided into 3 memory segments. 
 The data from the network switch is received by the
 net thread, which reassembles the data sequence as required by the $\tt xGPU$ code, 
 and stores the data in the so called GPU ring buffer. The GPU thread then processes the data in the buffer and computes the correlations. 
 After the GPU has finished the computation, it outputs the result to the so called CPU ring buffer. The CPU thread then 
 collects the result in the suitable format, which is then put in the so called disk ring buffer. Finally the disk thread saves the 
 data on the disk ring buffer to files on the storage system. 
In the pipeline, before one
thread starts to work, it will fill the ring buffer and alert the next
thread to prepare.
 For each ring buffer, 
 the up stream and down stream thread would write and read a different segment at the same time. 
 The size and number of the sub-buffers are selected to avoid data outflow.  
Hashpipe also provides a status buffer which can extract key-value pairs in each thread. This key-value is 
updated every running cycle. The status can be viewed using a GUI monitor that has been written in both Ruby and Python. 
The GPU CMAC is done by xGPU which is written by M.Clerk \citep{2013IJHPC..27..178C}. We also test the GPU performance of xGPU code with different number of antenna stations. For single GTX690 core, it will achieve peak performance 1.2TFLOPS when antenna station increase to 96, and keep stable with more inputs. We reach 42$\%$ of the official peak performance which is 2.8TFLOPS for a single core.\footnote{\url{https://www.geforce.com/hardware/desktop-gpus/geforce-gtx-690/performance}}

\subsection{Network and Data Packet.\label{sec:Local Network}}

The dish pathfinder array with a total of 32 inputs does not require the 10GbE network switch, as its data flow is still within the limit
of direct communication through 10GbE network. 
However, we designed the system with the flexibility and scalability, so that it could 
be extended to larger arrays. Indeed, we seek a design which can at least handle the 192 inputs of the cylinder array. 
Our design including 8 ROACH2s can handle a maximum of  256 inputs.
The design is also partially tested, though in the end it is not 
implemented in full. 

\begin{figure}
\centering
\includegraphics[width=0.6\textwidth]{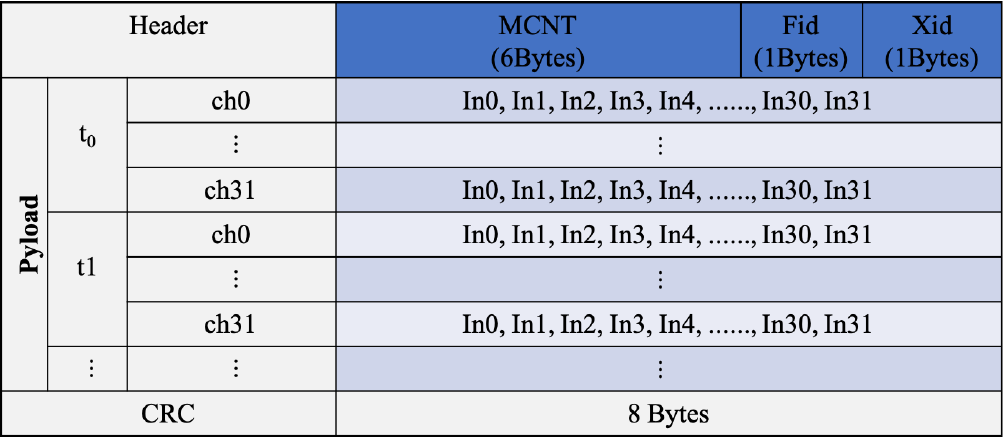}
\caption{Packet Data Format in our system. For 512 FFT length model, each time is 
from ch0 to ch15. \label{fig:dataformat}   }
\end{figure}

The data communication on the 10GbE is handled using the User Datagram Protocol (UDP). The  
data packet format is shown in Fig.\ref{fig:dataformat}, which includes the header, the content, 
and the cyclic redundancy checksum (CRC) for error correction. 
In the header,  after the UDP header, an application header is added 
to signify the packet properties. The application header is 8-bytes long, which is divided into three parts: 6 Bytes 
for packet counter MCNT, 1 byte for Fid, and 1 byte for Xid. The counter MCNT denotes the time sequence of the data, and the 
Hashpipe could detect packet loss by checking if there is any jump in the MCNT  sequence. The Fid is the identification of the ROACH2 node from which the packet is produced. The Xid denotes the GPU core where the package is sent, from which one could also know the frequency band  in the package.

When taking the correlations, the natural order is to take one receiver channel, then run cross-correlation with all channels (including 
itself as auto-correlation). The cross-correlations are computed
in time segments, with the data is in the format of [time, (fixed) channel, (run) input channel].
This is also shown in Fig.\ref{fig:dataformat}, where In0 means data from input 0. The data from the F-engine output is a complex 
number with a 4-bit real part and 4-bit imaginary part. Each packet contains $1/32$ of the whole frequency band, 
i.e. for 1024 frequency bins model, there are ${1024}/{32}=32$ frequency channels in each packet.
 In order to optimize the 10 GbE data transfer, we stack 8 time channels from the next time stamp at same frequency band in each packet. So the total size for one output packet from the F-engine is:
\begin{equation} 
N_{\rm chan}  \times N_{\rm input} \times 8 + \rm Header + CRC = 8208  (Bytes) .
\end{equation}

The data is transferred between the F-engine and X-engine via a 10GbE switch. In our system, each GPU node has four 
10GbE ports. As a result, for the 192-inputs we need a total of 
$4\times 6 $ ROACH2 nodes + $4 \times 8 $ GPU nodes = 56 ports on the 10GbE switch. 
We have a Mellanox$^{\rm TM}$ SX1024 Switch which has 48 ports of 10GbE and 12 ports of 40GbE. 
Furthermore, the 12 ports of 40 GbE can be split into $12\times4=48$ 10GbE ports. This would be sufficient for our purpose if we 
do build the correlator of the 192 inputs using this system.  

\begin{figure*}
 \centering
 \includegraphics[width=0.8\textwidth]{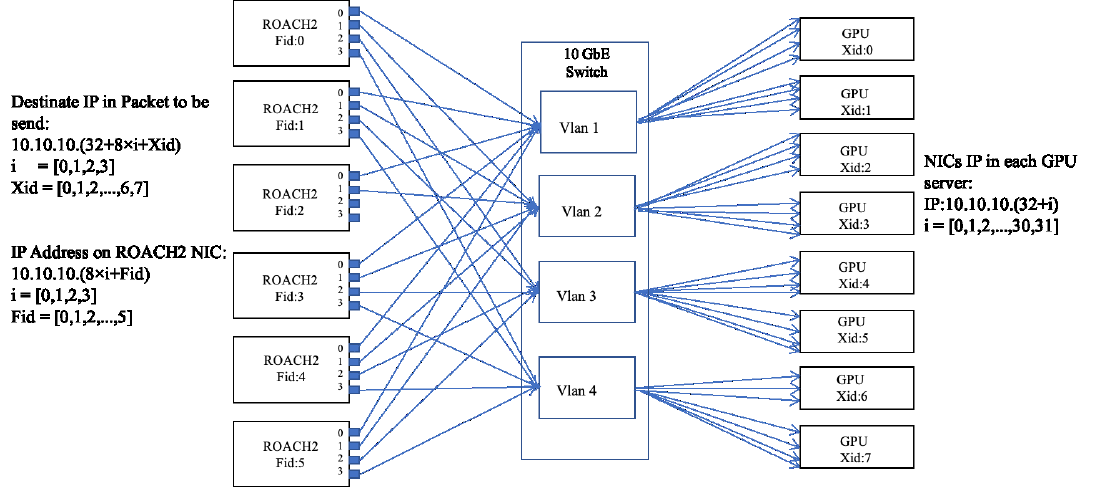}
 \caption{Network and IP assignment strategy in our system. The 10GbE Switch are divided into four Vlans.  \label{fig:network} }
\end{figure*}

In Fig.\ref{fig:network}, we show the IP assignment strategy in our network system. The 10 GbE ports on each ROACH2 node 
are denoted by index $i \in[0,1,2,3]$.  After the transpose, the output from the $i$th port of the ROACH2 board contains 
$\frac{BW}{4} \times i $ to $\frac{BW}{4} \times (i+1)$ frequency points between 8 packets going to different GPU cores. 
Taking port 0 for example, we will have 0\--256 frequency points. As each GPU core processes 32 frequency points, the data output 
from Port 0 must be sent to 8 GPU cores. Given that each GPU node has 4 GPU cores, only two nodes are required to receive the 
data, namely, node0 and node1.  Thus we can divide the ports of 10GbE switch into 4 subnets using  Virtual Local Area Network 
(VLAN), which will direct the data from port 0 of each ROACH2 board to GPU node0 and node1. The source of the 
packet received in the GPU core could be located from its Fid in the header, as discussed earlier.

\section{Tests and Experiments}\label{sec:experiment}

We have made a number of simple tests both in laboratory and on site to check the
performance of the correlator. 


In a frequency chirp test of the F-engine, we feed in a
monotone analog signal,  then after going through 
the F-engine we grab the packet using  Wireshark$^{\rm{TM}}$,  a software to view the UDP packet.
As expected, the F-engine produced the correct result.
We also tested the correlation of two identical signals with a delay. For the correlation of a signal of the form 
 $Ae^{i(2\pi ft + \phi)}$ and one with time delay $\tau$, the cross-correlation should be 
\begin{equation}
V = I_1^*\cdot I_2 = I e^{i2\pi f\tau}\label{eq:visibility}
\end{equation}
When the delay $\tau$ is a constant, the phase will be linear with frequency and have a slope of $k =2\pi \tau $. 
We use a power splitter to split the signal, and added  a  $L=7.5$m radio frequency 
cable in one of the signal path to simulate a time delay,
The result is consistent with a linear slope, with $\tau= L/ v$ where the signal speed is found to be $v=0.7c$, 
as expected for the signal speed in the RF cable used.  

We also did some test about the linearity of our correlator. By adjusting the input power of the GWN signal and the corresponding
output, we found good linearity in the range of -12dBm to 6dBm input power.
\begin{figure}
 \centering
 \includegraphics[width=0.5\textwidth]{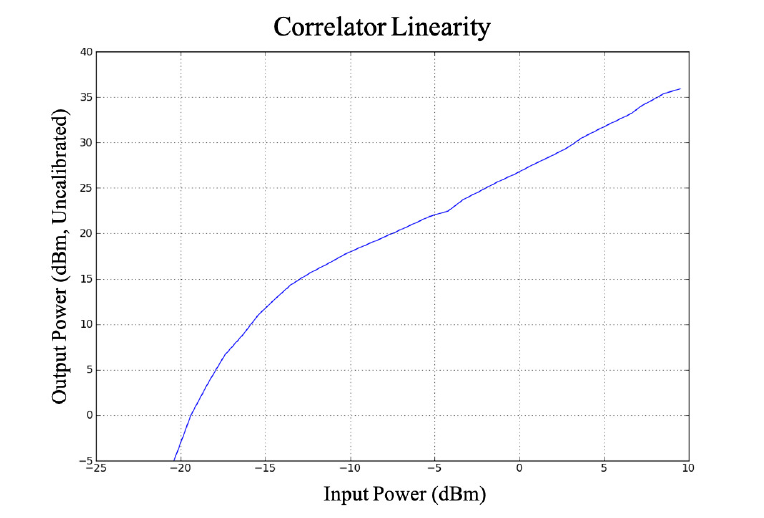}
\caption{linearity of This system, In the range of -12dBm to 6 dBm, this system has good linearity.\label{fig:linearity}}
\end{figure}

\begin{figure*}
 \centering
 \includegraphics[width=1\textwidth]{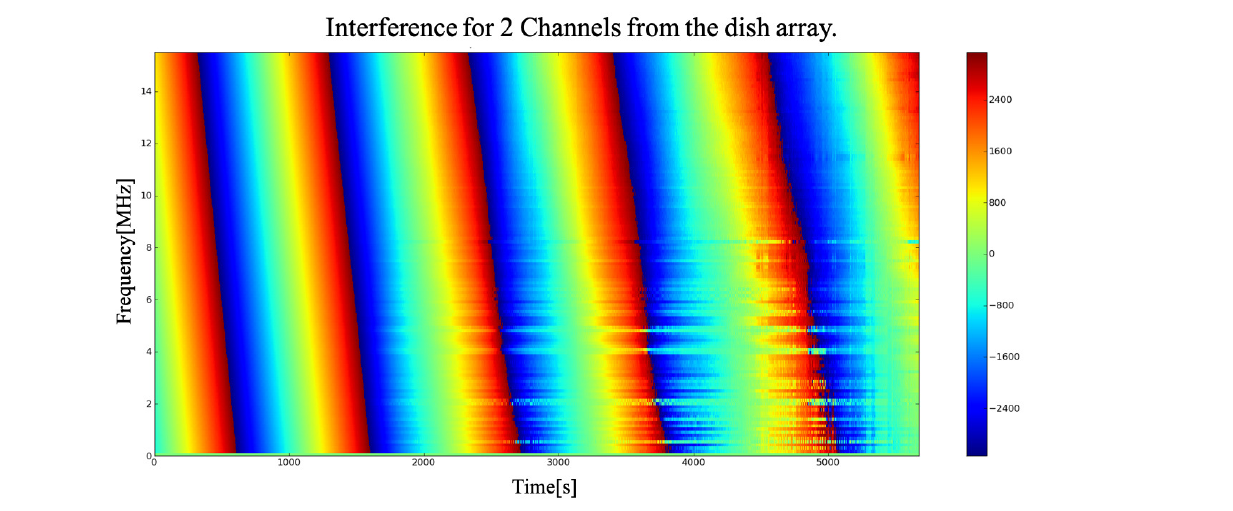}
\caption{Phase fringe between two telescope inputs by observing of sun.\label{fig:sun}}
\end{figure*}

Finally, the correlator was installed on the dish array and we observed bright radio sources such as the 
 Sun, Cygnus A, and other radio sources. Interference fringes can be clearly seen for the bright sources. 
Fig.\ref{fig:sun} shows the Visibility of the Sun through 2 inputs of the dish array during a period of 1.5 hours. 
The phase of visibility varied with frequency and time caused by Eq.\ref{eq:visibility}.  Note that the phase 
disturbance during the last 40 minutes is caused by external radio frequency interferences (RFI).

\section{Summary}\label{sec:summary}	
A correlator system based on the ROACH2-GPU framework is developed for the Tianlai Dish array with 32 inputs, and the design 
allows scalable expansion to larger array, e.g. the 192 inputs correlator. This correlator design is flexible and scalable. 
Two FFT lengths, 1024 and 512, are implemented. In the correlator system, different ROACH2 boards and X-engine are running the 
same F-engine gateware and X-engine sofware with different parameter. The X-engine consists of $\tt xGPU$ computation core and Hashpipe data flow 
manage system. The hardware in our hand is adequate for the 32 inputs correlator, but for 192 inputs correlator, it can only handle 
$1/8$ bandwidth right now. The only thing that needs modification in the software is  to change some parameters such as frequency band 
configuration for the F-engine and X-engine. Furthermore,  the ROACH2-SWITCH-GPU framework can also be used for different purposes 
at the same time. The switch has a broadcast mechanism,  that we could put in other backend, e.g. an   FRB search 
backend to the system by extending the switch system and adding some FRB nodes.  
We also did some experiments with our instrument, and the 32-inputs correlator system worked normally.  Given the 
additional hardware, it can also be extended to the 192-input system, or even to larger systems in the future.

\section*{Acknowledgements}
We thank the support from the CASPER community for making available the hardware and software they developed, 
and offering helps when we encountered problems. Chenhui Niu acknowledges the China Scholarship Council for providing support
of his visit to the CASPER group in UC Berkeley. The Tianlai project has been supported by the 
 the  Repair and Procurement Program of the Chinese Academy of Science, the NSFC grant 11473044, 11633004, 
MoST grant 2016YFE0100300 and 2012AA121701, the  NSFC grant 11473044, 11633004, 11761141012, 
and the CAS Frontier Science Key Project QYZDJ-SSW-SLH017.

\end{document}